# Mixed Reality Interface for Digital Twin of Plant Factory

Byunghyun Ban
Imagination Garden Inc.
Andong, Republic of Korea
halfbottle@sangsang.farm

*Abstract*—An easier and intuitive interface architecture is necessary for digital twin of plant factory. I suggest an immersive and interactive mixed reality interface for digital twin models of smart farming, for remote work rather than simulation of components. The environment is constructed with UI display and a streaming background scene, which is a real time scene taken from camera device located in the plant factory, processed with deformable neural radiance fields. User can monitor and control the remote plant factory facilities with HMD or 2D display based mixed reality environment. This paper also introduces detailed concept and describes the system architecture to implement suggested mixed reality interface.

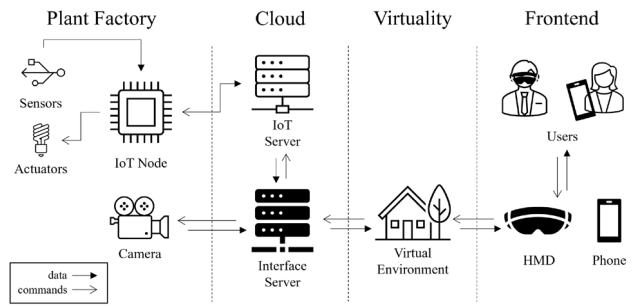

Fig. 1. Summary of system architecture

## I. Introduction

### A. Digital Twin for Plant Factory

Previously, both smart farming solutions for IoT based automation [1-5] and precise environment analysis [6-11] were described as computer-based processing solutions for real-world only. However, applying digital twin models for agricultural purpose is now considered as a common inspiration among many smart farm engineers [16-20]. Digital twin methods provide precise analysis of current state and prediction of future state of industrial systems [12-15].

Hwang, et al. [16] and Alves, et al. [17] suggested a simulation model which can describe the state of real-world farm, including not only just common sensor-measured values but also the growth state of plants, and can receive control commands for actuators. As their approaches are applicable to open-field farms only, advanced solutions for indoor farm are required for most smart farm industry. Ko, et al. [18] and Monterio, et al. [19] has provided a specific service architecture for implementation of digital twin for indoor vertical farming systems, applied with existing IoT architecture. And Jeong, et al. [20] has shown that digital twin model for indoor plant factory can perform very precise real time analysis of environment state in high resolution.

Recent approaches provide useful blueprints for plant factory digital twin model. However, their 3D-rendered graphic-based model has many limitations. First, their methods always require pre-built 3D virtual farm rendering models for operation. It is only affordable for a small environment, and not appropriate for industrial-scaled plant factories. Secondly, once a 3D digital twin model is built, the users cannot move the facilities of real-word farms because it will lead to inconsistence between the twin. Also, a digital twin model should be easy to understand for old-aged users.

Therefore, I suggest a mixed reality environment with 3D scene reconstruction technologies to provide an immersive and intuitive digital twin environment.

### B. 3D scene processing and reconstruction

3D scene processing method with pictures taken from a rotating camera or multi angle cameras has been very popular topic in computer vision processing area. Many researchers have suggested noble methods to overlap discrete images to build a continuous scene of 3D environment [21-22]. So various technologies or products for 360 ° scene are available now. Even some engineers now focus on how to reduce the camera cost [23] rather than image quality enhancement.

Thanks to NeRF [24], a single camera with plain angle of view can construct a 360 ° scene. NeRF model can generate the scenes between discrete pictures taken with single camera with various angle. There also are several applications of the original NeRF model such as depth-supervised model [25] and 3D scene generative model [26]. Even Nerfies model [27] can reconstruct deformable radiance fields to process a scene from pictures of moving object, or from pictures taken by moving camera. Therefore, a 360 ° scene can be generated with just one rotating camera, to provide an immersive VR-like environment.

## II. Mixed reality

### A. Concept

While applying virtual elements such as buttons or indicators on real-world data is considered as augmented reality, augmented virtuality is its opposite concept: applying real-world element on virtual environment. However, in the proposed method, the real-word data is taken from remote farms and processed into 3D VR-like scene. Therefore, I think that the concept of proposed method should be categorized as mixed reality technology.

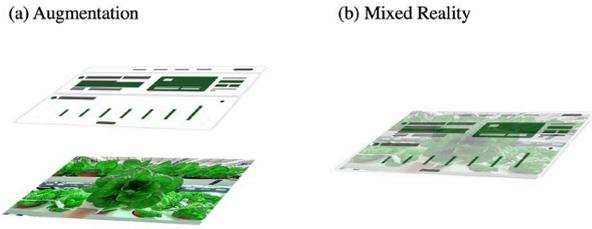

Fig. 2. Concept of mixed reality for plant factory digital twin

Fig. 2 shows the main concept of the interface environment. The main element of digital twin model is the virtual components such as control panel to monitor and control the remote smart farm environment

### B. Interface

Users can access the interface system both via HMD or 2D display devices. The interface provides an immersive and intuitive scene synthesized with virtual interfaces for controlling and monitoring of farms, and real time streaming video taken from remote plant factory. User can interact with the environment with FPS game-like experiences. The interface continuously provides information of interactable object and options of additional actions for manipulation of active state of actuators.

The interface directly communicates with IoT servers to send and receive information and commands. Proposed method is fully compatible with previously installed IoT system.

## III. METHODS

### A. System Architecture

Fig. 1 describes brief architecture of the system. Additional components for security options or data storage efficiency are recommended but not described on the illustration.

The sensors, actuators, IoT Node, and IoT server are not different from previously installed common-type IoT infrastructures. IoT server provides communication protocols such as rest APIs, which was previously provided with frontend applications. In this method, APIs of IoT servers are not directly applied on the frontend side. The interface server calls APIs of IoT server to draw data and provide commands from the users.

The interface server also receives real time visual data from the cameras which are located in the plant factory. It also processes the images with 3D scene synthesis algorithms in order to construct a VR-like 3D environment stream. Then the interface server mixes virtual UI with 3D streaming scene to produce a mixed reality environment.

### B. User Interface Devices

Users can approach to the mixed reality environment with interface devices such as HMD or phones to monitor remote plant factory and send commands.

### C. Command Stream

Users can make 3 kinds of command streams.

#### 1) Commands for IoT Server

Commands for IoT server is just same as the user commands from the frontend interface of traditional IoT. However, the commands are not directly sent to IoT server. The user interface device first receives the commands, and it send them to the interface server. Finally, the interface server calls API of IoT servers to send user's commands.

Users can make not only commands for more information of environment state of remote plant factory, but also commands for manipulation of actuators' activation conditions, activation periods, or just activation state.

#### 2) Commands for Interface Server

User can provide commands for the interface server too, via user interface devices. Users can request the interface server for interface pop-up, camera movements, object selection, and any other commands for virtual environment control.

#### 3) Commands for User Interface Devices.

User also commands the interface devices for view scope angle change, zoom in and out, cursor indicator control, and any others for UI operation and experiences.

### D. Scene Acquirement

Camera devices should acquire real time scene for the proposed method. Both stationary cameras and self-driving-like moving camera devices can be applied for mixed reality environment. I recommend to utilize 360° camera attached on a self-driving device, but a rotating single normal camera is compatible with this system if deformable neural radiance field [27] based image processing algorithms are available.

### E. 3D Scene Synthesis

2D pictures taken from stationary camera data can be processed with traditional panoramic scene synthesis algorithms. However, if the camera device moves around the plant factory, I highly recommend applying NeRF-like applications which can remove artifacts driven by the movement of camera [27-29] or artifacts driven by LED devices [30-31] to build 3D scenes, Because of dense and irregular shapes of leaves and countless LED chips around the plant factory. I used Nerfies [27] for proposed method.

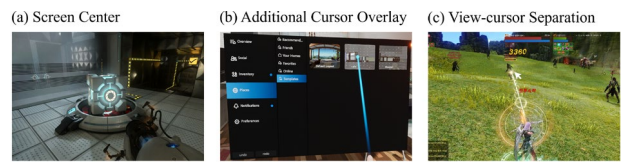

Fig. 3. Common interaction interfaces among FPS games

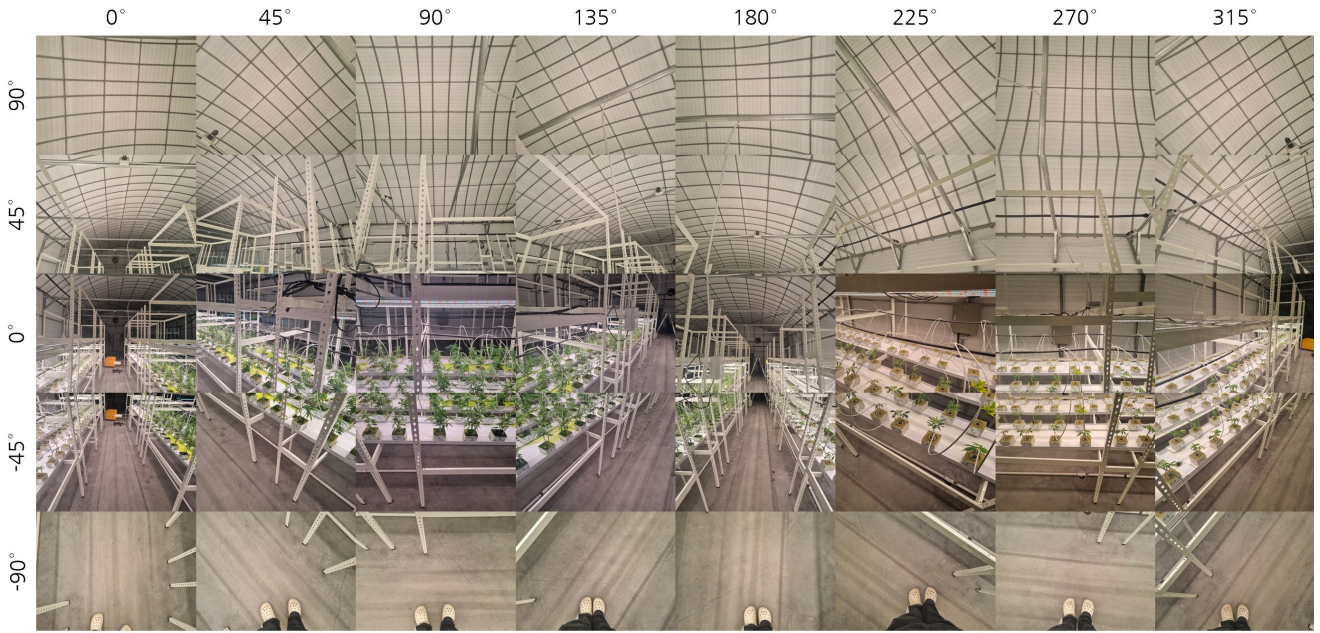

Fig. 4. 360° picture taken by a rotating single camera, around the plant factory

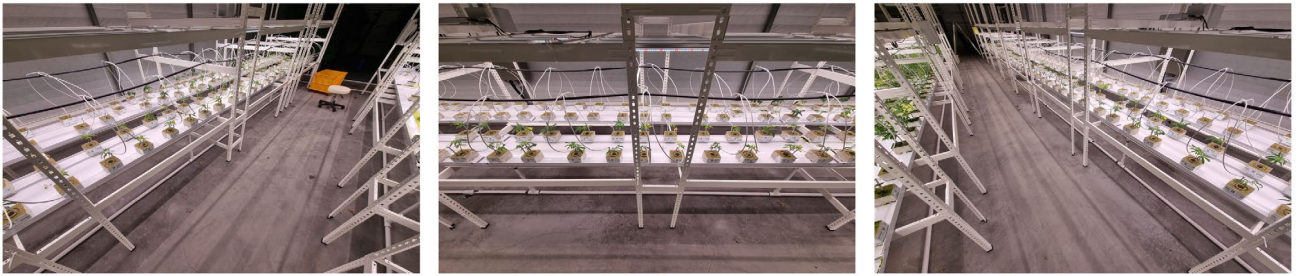

Fig. 5. Synthesized scenes after postprocessing

### F. Interaction methods with Objects

As the proposed virtual environment displays the streaming scene of plant factory as a first person view, the interface structure for user-object interaction is very similar to FPS games. The user interface devices can provide 3 different types of interaction settings which are illustrated on Fig. 3.

Screen center method [32] are recommended for desktop computer users. The interface always selects an interactable object which is on the center of screen. So the user just can quickly change the angle of view with mouse, and simply move the view for object selection.

Additional cursor overlay method [33] is highly recommend for VR-ready devices such as HMD. User holds additional remote controller on the hand to control the indicator. While user can change the view by rotation of the head, the controller also displays a virtual indicator on the scene.

Touch screen based displays such as phones or tablet computers may provide best experience with view-cursor separation method [34]. User needs to use their fingers to change the view angle and zoom level. So complete separation between screen and cursor is recommended for comfortable UX. Users can just touch any interactable object displayed on the screen.

### G. Interpolation of Sensor Values

There are limited numbers of sensors for environment monitoring. Therefore, the IoT server predicts missing values of environment states by interpolation among the sensor values. The IoT server stores (x, y, z) coordinates of sensors. And when the interface server requires the state value of (X, Y, Z), which is the location where the user interface device has received by users, the IoT server calculates the predicted values from installed sensor's locations and values.

## IV. RESULT

### A. View Synthesis

A set of 360° camera scene was taken with rotating single camera around the plant factory, taking pictures at every 45°. Then the pretrained Nerfies [27] model processed the scenes to generate new scenes. Fig. 4 displays the input images taken from medical hemp factory of Imagination Garden Inc. in

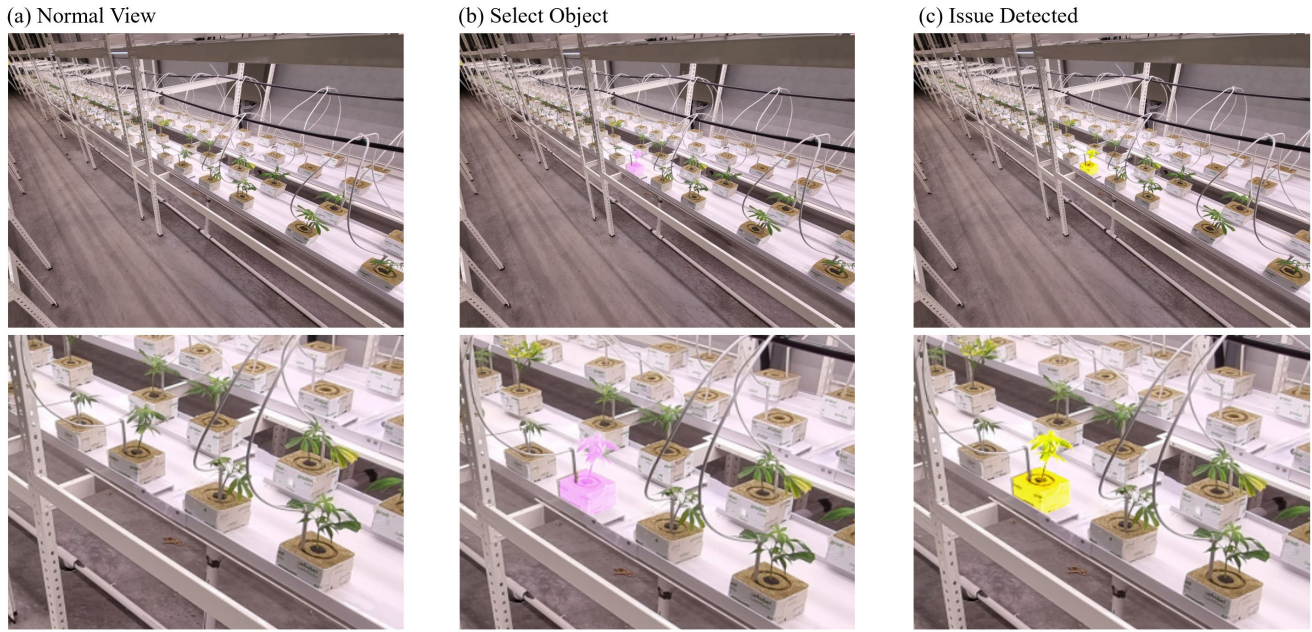

Fig. 6. Object highlighting

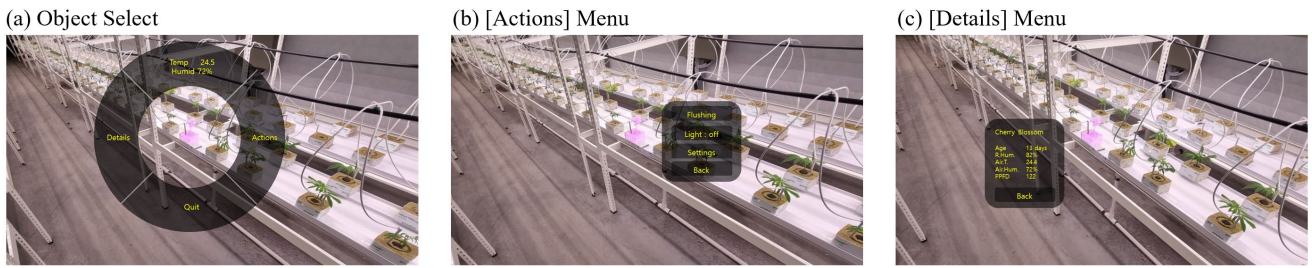

Fig. 7. User-object interaction interface

Andong-si, Republic of Korea. And Fig. 5 was the synthesized scenes after postprocessing done.

*B. Interaction with Objects*

Fig. 6 illustrates object highlighting scenarios for user-object interaction. (a) shows normal state view. (b) shows an object activated by user's action such as click or indicator overlay. (c) describes an object with issues which users should check up.

*1) Interaction Interface*

Interaction UI examples are described on Fig. 7. When user activate the interactable object, summarized environment information and buttons for [Actions] or [Details] pop up. [Action] menu displays command options for actuators which the IoT servers can receive. [Details] menu shows detailed information of the object, and interpolated state variables of the location where the object is.

*2) Issue Report*

The environment shows issue state with color change. The severity of issue is expressed with color change.

V. DISCUSSION

As 45° angle was too wide to build a radiance field, small amounts of pictures showed with satisfying quality. So, I think a 360° camera is still needed to provide a perfect scene, rather than processing scenes taken from a rotating camera. Also, further development for real time streaming and user-interface interaction are necessary for a commercial level application.

VI. CONCLUSION

User can experience 3D-like mixed reality environment with proposed system with available interface devices. The visual information from camera and sensory information from IoT nodes are processed in an intuitive way and provided to users in easy UI. Also user can make control commands of virtual objects, which lead to actuator state change. Therefore, the proposed method provides a digital twin for remote control and monitoring of a remote plant factory.

ACKNOWLEDGMENT

I would like to express my deep gratitude to Donghun Ryu for taking so many pictures of the plant factory.